\documentclass{gen-p-l}

\usepackage{amsmath}
\usepackage{amssymb}
\usepackage{amsthm}
\usepackage{latexsym}

\def\R{{\bf R}}

\def\E{{\mathcal E}}

\def\x{{\vec{x}}}
\def\y{{\vec{y}}}
\def\k{{\vec{k}}}

\def\mfr#1/#2{\hbox{$\frac{{#1}}{{#2}}$}}
\def\uprho{\raise1pt\hbox{$\rho$}}
\def\upchi{\raise1pt\hbox{$\chi$}}
\def\dlambda{\lower1pt\hbox{$\lambda$}}
\newcommand{\xij}{|x_i-x_j|}
\def\R{{\bf R}}

\def\E{{\mathcal E}}

\def\x{{\vec{x}}}

\def\mfr#1/#2{\hbox{$\frac{{#1}}{{#2}}$}}
\def\uprho{\raise1pt\hbox{$\rho$}}
\def\upchi{\raise1pt\hbox{$\chi$}}
\def\dlambda{\lower1pt\hbox{$\lambda$}}

\newtheorem{theorem}{Theorem}[section]
\newtheorem{lemma}[theorem]{Lemma}

\theoremstyle{definition}

\theoremstyle{remark}

\numberwithin{equation}{section}

\newtheorem{corollary}[theorem]{Corollary}

\begin{document}

\title{THE BOSE GAS: A SUBTLE MANY-BODY PROBLEM}

% Remove or comment out any unused author tags.
% author one information
%\author{Elliott H.\ Lieb}
%\address{Departments of Physics and Mathematics, Jadwin
%Hall,
%Princeton University,\newline P.~O.~Box 708, Princeton, New Jersey
%  08544, USA}
%%\curraddr{}
%\email{lieb@princeton.edu}
%\thanks{The author was supported in part by
%U.S. National Science Foundation grant PHY 98-20650 A01. \\
%\copyright 2000 by the author. Reproduction of this work, in its
%entirety, by any means, is permitted for non-commercial purposes.

\author{Elliott H.\ Lieb}
\address{Departments of Mathematics and Physics, Princeton University,
Princeton, New Jersey 08544-0708}
\email{lieb@princeton.edu}

\thanks{Supported in part by NSF Grant PHY-98 20650 A01. }

\thanks{\copyright 2000 by the author. Reproduction of this article, in its
entirety, by any means, is permitted for non-commercial purposes.}

\thanks{For the proceedings of the {\it XIII International Congress of
Mathematical Physics}, London, July 18-24, 2000.}

\date{August 19, 2000 }

\dedicatory{  }

\begin{abstract}
Now that the properties of  the ground state of quantum-mechanical
many-body systems (bosons) at low density, $\rho$, can be examined
experimentally it is appropriate to revisit some of the formulas deduced
by many authors 4-5 decades ago.  One of these is that the leading term in
the energy/particle is $4\pi a \rho$ where $a$ is the scattering length
of the 2-body potential.  Owing to the delicate and peculiar nature of
bosonic correlations (such as the strange $N^{7/5}$ law for charged
bosons), four decades of research failed to establish
this plausible formula rigorously. The only previous lower bound for
the energy was found by Dyson in 1957, but it was 14 times too small.
The correct asymptotic formula has recently been obtained jointly
with J. Yngvason and this work will be presented. The reason behind
the mathematical difficulties will be emphasized. A different formula,
postulated as late as 1971 by Schick, holds in two-dimensions and this,
too, will be shown to be correct.  Another problem of great interest
is the existence of Bose-Einstein condensation, and what little is
known about this rigorously will also be discussed.  With the aid of
the methodology developed to prove the lower bound for the homogeneous
gas, two other problems have been successfully addressed. One is the
proof (with Yngvason and Seiringer) that the Gross-Pitaevskii equation
correctly describes the ground state in the `traps' actually used in
the experiments.  The other is a very recent proof (with Solovej) that
Foldy's 1961 theory of a high density gas of charged particles correctly
describes its ground state energy.  \end{abstract}

\maketitle

\section{INTRODUCTION}\label{intro}

Schr\"odinger's equation of 1926 defined a new mechanics whose
Hamiltonian is based on classical mechanics, but whose consequences are
sometimes non-intuitive from the classical point of view. One of the most
extreme cases is the behavior of the ground (= lowest energy) state of a
many-body system of particles. Since the ground state function
$\Psi(x_1,...,x_N)$ is automatically symmetric in the coordinates
$x_j\in R^3$ of the $N$ particles, we are dealing necessarily with {\it
`bosons'.} If we imposed the Pauli exclusion~principle (antisymmetry)
instead, apppropriate for electrons,  the outcome would look much more
natural and, oddly, more classical. Indeed, the Pauli principle is {\it
essential} for understanding the stability of the ordinary matter  that
surrounds us.

Recent experiments have confirmed some of the bizarre properties of
bosons close to their ground state, but the theoretical ideas go back to
the 1940's - 1960's. The first sophisticated analysis of a gas or liquid
of {\it interacting} bosons is due to Bogolubov in 1947. His
approximate theory as amplified by others, is supposed to be exact in
certain limiting cases, and some of those cases have now been verified
rigorously (for the ground state energy) -- 3 or 4 decades after they were
proposed.

The discussion will center around three main topics.

\bigskip

\noindent
1.\textsf{ The dilute, repulsive Bose gas at low
density (2D and 3D)} (with Jakob Yngvason).

\medskip

\noindent 
2.\textsf{ Repulsive bosons in a trap (as used in recent
experiments) and the `Gross-Pitaevskii equation} (with Robert Seiringer and
Jakob Yngvason).

\medskip

\noindent
3.\textsf{ Foldy's `jellium' model of charged particles in 
a neutralizing background} (with Jan Philip Solovej)

\medskip

The discussion below of topic 1 is taken from \cite{LY1998} (in
3-dimensions) and \cite{LY2d} (in 2-dimensions). That for topic 2 is taken
from \cite{LSY1999}  (in 3-dimensions) and \cite{LSY2d} (in 2-dimensions).
Topic 3 is from \cite{LS}. See also \cite{LYbham, LSYdoeb}.

Topic 1 (3-dimensions) was the starting point and contains essential
ideas. It is explained here in some detail and is taken, with minor
modifications (and corrections), from \cite{LYbham}. In terms of 
technical complexity, however, the third topic is the most
involved and can be treated here only very briefly.

\bigskip

\section{THE DILUTE BOSE GAS IN 3D} \label{sect3d}

We consider the Hamiltonian for $N$ bosons of mass 
$m$ enclosed in a cubic box $\Lambda$ of side length $L$ and interacting by a
spherically symmetric pair potential 
$v(|\x_i - \x_j|)$:
\begin{equation}\label{ham}
H_{N} = - \mu\sum_{i=1}^{N} \Delta_i + 
\sum_{1 \leq i < j \leq N} v(|\x_i - \x_j|).
\end{equation}
Here  $\x_i\in\mathbb R^3$, $i=1,\dots,N$ are the positions of the particles,
$\Delta_i$ the Laplacian with respect to $\x_{i}$,
and we have denoted ${\hbar^2}/{ 2m}$ by $\mu$ for short. (By choosing 
suitable units $\mu$ could, of course, be eliminated, but we want to keep 
track 
of the dependence of the energy on  Planck's constant and the mass.) The 
Hamiltonian
(\ref{ham}) operates on {\it symmetric} wave functions in $L^2(\Lambda^{N}, 
d\x_1\cdots d\x_N)$ as is appropriate for bosons. The interaction 
potential will be assumed to be {\it nonnegative} and to decrease 
faster than $1/r^3$ at infinity.
 
We are interested in the ground state energy $E_{0}(N,L)$ of (\ref{ham}) in 
the 
{\it thermodynamic limit} when $N$ and $L$ tend to infinity with the 
density $\rho=N/L^3$ fixed. The energy per particle in this limit
\begin{equation} e_{0}(\rho)=\lim_{L\to\infty}E_{0}(\rho L^3,L)/(\rho 
L^3).\end{equation}
Our results about $e_{0}(\rho)$ are based on estimates on
$E_{0}(N,L)$
for finite $N$ and $L$, which are important, e.g., for the considerations of 
inhomogeneous systems in \cite{LSY1999}. 
To define  $E_{0}(N,L)$ precisely one 
must specify the boundary conditions. These should not matter for the
thermodynamic limit. 
To be on the safe side we use Neumann boundary conditions for the 
lower bound, and Dirichlet boundary conditions for the upper bound 
since these lead, respectively, to the lowest and the highest energies.

For experiments with dilute gases the {\it low density asymptotics} of 
$e_{0}(\rho)$ is of importance. Low density means here that the mean 
interparticle distance, $\rho^{-1/3}$ is much larger than the 
{\it scattering length} $a$ of the potential, 
%%%% ADDITION
which is defined as
follows. The zero energy scattering Schroedinger equation 
\begin{equation}\label{3dscatteq}
-2\mu \Delta \psi + v(r) \psi =0
\end{equation}
has a solution  of the form, asymptotically as $r\to \infty$ (or for all
$r>R_0$ if $v(r)=0$ for $ r>R_0$),
\begin{equation}\label{3dscattlength}
\psi(x) = 1-a/|x|
\end{equation}
This is the same as 
%%%%%END ADDITION
\begin{equation}a=\lim_{r\to\infty}r-\frac{u_{0}(r)}{u_{0}'(r)},\end{equation}
where $u_{0}$ solves the zero energy scattering equation,
\begin{equation}\label{scatteq}
-2\mu u_{0}^{\prime\prime}(r)+ v(r) u_{0}(r)=0
\end{equation}
with $u_{0}(0)=0$. (The factor $2$ in (\ref{scatteq}) comes from 
the reduced mass of the two particle problem.) 

An important special case is the hard core potential $v(r)= \infty$ if
$r<a$ and  $v(r)= 0$ otherwise. Then the scattering length $a$ and the
radius $a$ are the same.

Our main result is a 
rigorous proof of the formula
\begin{equation} e_{0}(\rho)\approx4\pi\mu\rho a\end{equation}
for $\rho a^3\ll 1$, more precisely of
\begin{theorem}[\textbf{Low density limit of the ground state energy}]
\begin{equation}\label{basic}
\lim_{\rho a^3\to 0}\frac {e_{0}(\rho)}{4\pi\mu\rho a}=1.	
\end{equation}	
\end{theorem}
This formula is independent of the boundary conditions used for the 
definition of $e_{0}(\rho)$.

The genesis of an understanding of $e_{0}(\rho)$ was the pioneering 
work \cite{BO} of Bogolubov, and in the 50's and early 60's several
derivations of (\ref{basic}) were presented \cite{Lee-Huang-YangEtc}, 
\cite{Lieb63}, even including higher order terms:
\begin{equation}\frac{e_{0}(\rho)}{4\pi\mu\rho a}=
1+\mfr{128}/{15\sqrt \pi}(\rho a^3)^{1/2}
+8\left(\mfr{4\pi}/{3}-\sqrt 3\right)(\rho a^3)\log (\rho a^3)
+O(\rho a^3)
\end{equation}
These early developments are reviewed in \cite{EL2}. They all rely 
on some special assumptions about the ground state that have never been 
proved, or on the selection of special terms from a perturbation series 
which likely diverges. The only rigorous estimates of this period were 
established by Dyson, who derived the following bounds in 1957 for a 
gas of hard spheres \cite{dyson}: 
\begin{equation} \frac1{10\sqrt 2} \leq
	\frac{e_{0}(\rho)}{ 4\pi\mu\rho a}\leq\frac{1+2 Y^{1/3}}{ 
(1-Y^{1/3})^2}
\end{equation}
with $Y=4\pi\rho a^3/3$. While the upper bound has the asymptotically 
correct form, the lower bound is off the mark by a factor of about 1/14.
But for about 40 years this was the best lower bound available!

Under the assumption that (\ref{basic}) is a correct asymptotic formula
for the energy, we see at once that understanding it physically, much
less proving it, is not a simple matter. Initially, the problem prsents
us with two lengths, $a \ll \rho^{-1/3}$ at low density. However,
(\ref{basic}) presents us with another length generated by the solution
to the problem. This length is the de Broglie wavelength, or
`uncertainty principle' length 
$$
\ell_c = (\rho a)^{-1/2} .
$$
The reason for saying that $\ell_c$ is the de Broglie wavelength is that 
in the hard core case all the energy is kinetic (the hard core just
imposes a $\psi =0$ boundary condition whenever the sdistance between
two particles is less than $a$). By the uncertainty principle, the
kinetic energy is proportional to an inverse length squared, namely
$\ell_c$. 
We then have the relation (since $\rho a ^3$ is small)
$$
a \ll \rho^{-1/3}\ll \ell_c  
$$
which implies, physically, that it is impossible to localize the
particles relative to each other (even though $\rho$ is small)
bosons in their ground state are therefore `smeared out' over distances 
large compared to the mean particle distance and their individuality 
is entirely lost. They cannot be localized with respect to each other
without changing the kinetic energy enormously.

Fermions, on the other hand, prefer to sit in 
`private rooms', i.e., $\ell_{c}$ is never bigger than $\rho^{-1/3}$
by a fixed factor. 
In this respect the quantum nature of bosons is much more pronounced 
than for fermions.

Since (\ref{basic}) is a basic result about the Bose gas it is clearly
important to derive it rigorously and in reasonable generality, in
particular for more general cases than hard spheres.  The question
immediately arises for which interaction potentials one may expect it
to be true. A notable fact is that it {\it not true for all} $v$ with
$a>0$, since there are two body potentials with positive scattering length
that allow many body bound states \cite{BA}. Our proof, presented in the
sequel,  works for nonnegative $v$, but we conjecture that (\ref{basic})
holds if $a>0$ and $v$ has no $N$-body bound states for any $N$. The
lower bound is, of course, the hardest part, but the upper bound is not
altogether trivial either.

%%%%%%%%
Before we start with the estimates a simple computation and some 
heuristics may be helpful to make 
(\ref{basic}) plausible and motivate the formal proofs.

With  $u_{0}$ the scattering solution and 
$f_{0}(r)=u_{0}(r)/r$,
partial integration gives
\begin{eqnarray}\label{partint}
\int_{|\x|\leq R}\{2\mu|\nabla f_{0}|^2+v|f_{0}|^2\}d\x&=&
4\pi\int_{0}^{R}\{2\mu[u_{0}'(r)-(u_{0}(r)/r)]^2+v(r)|u_{0}(r)]^2\}dr
\nonumber\\&=&
8\pi\mu a |u_{0}(R)|^2/R^2\to 8\pi\mu a\quad\mbox{\rm for $R\to\infty$},
\end{eqnarray}
if $u_{0}$ is normalized so that $f_{0}(R)\to 1$ as $R\to\infty$.
Moreover, for positive interaction potentials the scattering solution 
minimizes 
the quadratic form in (\ref{partint}) for each $R$ with 
$u_{0}(0)=0$ and $u_{0}(R)$ fixed as boundary conditions. Hence the energy 
$E_{0}(2,L)$ of two 
particles in a large box, i.e., $L\gg 
a$, is approximately $8\pi\mu a/L^3$. If the gas is sufficiently 
dilute it is not unreasonable to expect that the energy is essentially 
a sum of all such two particle contributions. Since there are 
$N(N-1)/2$ pairs, we are thus lead to $E_{0}(N,L)\approx 4\pi\mu a 
N(N-1)/L^3$, which gives (\ref{basic}) in the thermodynamic limit.

This simple heuristics is far from a rigorous proof, however, 
especially for the lower bound. In fact, it is  rather remarkable that 
the same asymptotic formula holds both for `soft' interaction 
potentials, where perturbation theory can be expected to be a good 
approximation, and potentials like hard spheres where this is not so.
In the former case the ground state is approximately the constant 
function and the energy is {\it mostly potential}:
According to perturbation theory
$E_{0}(N,L)\approx  N(N-1)/(2 L^3)\int v(|\x|)d\x$. In particular it is {\it 
independent of} $\mu$, i.e. of Planck's constant and mass. Since, 
however, $\int v(|\x|)d\x$ is the first Born approximation to $8\pi\mu 
a$ (note that $a$ depends on $\mu$!), this is not in conflict with 
(\ref{basic}).
For `hard' potentials on the other hand, the ground state is {\it 
highly correlated}, i.e., it is far from being a product of single 
particle states. The energy is here {\it mostly kinetic}, because the 
wave function is very small where the potential is large. These two 
quite different regimes, the potential energy dominated one and the 
kinetic energy dominated one, cannot be distinguished by the low 
density asymptotics of the energy. Whether they behave 
differently with respect to other phenomena, e.g., Bose-Einstein 
condensation, is not known at present.

Bogolubov's analysis \cite{BO} presupposes the existence of Bose-Einstein
condensation. Nevertheless, it is correct (for the energy) for the
one-dimensional delta-function Bose gas \cite{LL}, despite the fact
that there is (presumably) no condensation in that case. It turns
out that BE condensation is not really needed in order to understand
the energy.  As we shall see,  `global' condensation can be replaced
by a `local' condensation on boxes whose size is independent of
$L$. It is this crucial understanding that enables us to prove Theorem
1.1 without having to decide about BE condensation.

An important idea of Dyson was to transform the hard sphere 
potential into a soft potential at the cost of sacrificing the 
kinetic energy, i.e., effectively to move from one 
regime to the other. We shall make use of this idea in our proof
of the lower bound below. But first we discuss the simpler upper 
bound, which relies on other ideas from Dyson's beautiful paper \cite{dyson}.

%%%%%%%%%%%
\subsection{UPPER BOUND}

The following generalization of Dyson's upper bound holds 
\cite{LSY1999}, \cite{S1999}:
\begin{theorem}[\textbf{Upper bound}] Define $\rho_{1}=(N-1)/L^3$ and 
$b=(4\pi\rho_{1}/3)^{-1/3}$. For nonnegative potentials $v$, and $b>a$ 
the ground state energy of (\ref{ham}) with periodic boundary conditions 
satisfies
\begin{equation}\label{upperbound}
E_{0}(N,L)/N\leq 4\pi \mu \rho_{1}a\frac{1-\frac{a} 
{b}+\left(\frac{a} 
{b}\right)^2+\frac12\left(\frac{a} 
{b}\right)^3}{\left(1-\frac{a} 
{b}\right)^8}.	
\end{equation}
For Dirichlet boundary conditions the estimate holds with ${\rm 
(const.)}/L^2$ added to the right side.
Thus in the thermodynamic limit and for all boundary conditions
\begin{equation}
\frac{e_{0}(\rho)}{4\pi\mu\rho a}\leq\frac{1-Y^{1/3}+Y^{2/3}-\mfr1/2Y}
{(1-Y^{1/3})^8}.
\end{equation}
provided $Y=4\pi\rho a^3/3<1$.
\end{theorem}
{\it Remark.} The bound (\ref{upperbound}) holds for potentials 
with infinite range, provided $b>a$. For potentials of finite range 
$R_{0}$ it can be improved for $b>R_{0}$ to
\begin{equation}\label{upperbound2}
E_{0}(N,L)/N\leq 4\pi \mu \rho_{1}a\frac{1-\left(\frac{a} 
{b}\right)^2+\frac12\left(\frac{a} 
{b}\right)^3}{\left(1-\frac{a} 
{b}\right)^4}.	
\end{equation}

{\it Proof.}
We first remark that the expectation value of (\ref{ham}) with any 
trial wave function gives an upper bound to the bosonic ground state 
energy, even if the trial function is not symmetric under permutations 
of the variables.  The reason is that an absolute ground state of the 
elliptic differential operator (\ref{ham}) (i.e., a ground state 
without symmetry requirement) is a nonnegative function which can be 
be symmetrized without changing the energy because (\ref{ham}) is 
symmetric under permutations.  In other words, the absolute ground 
state energy is the same as the bosonic ground state energy.

Following \cite{dyson} we choose a trial function of
the following form
\begin{equation}\label{wave}
\Psi(x_1,\dots,x_N)=F_1(x_{1}) \cdot F_2(x_{1},x_{2}) \cdots 
F_N(x_{1},\dots,x_{N}).
\end{equation}
More specifically, $F_{1}\equiv 1$ and $F_{i}$
depends only on the distance of $x_{i}$ to its nearest neighbor among 
the the points  $x_1,\dots ,x_{i-1}$ (taking the periodic boundary 
into account):
\begin{equation}\label{form}
F_i(x_1,\dots,x_i)=f(t_i), 
\quad t_i=\min\left(\xij,j=1,\dots, 
i-1\right),
\end{equation}
with a function $f$ satisfying  
\begin{equation}0\leq 
f\leq 1, \quad f'\geq 0.
\end{equation}
The intuition behind the ansatz (\ref{wave}) is that the particles are 
inserted into the system one at the time, taking into account the 
particles previously inserted. While such a wave function cannot 
reproduce all correlations present in the true ground state, it turns 
out to capture the leading term in the energy for dilute gases. 
The form (\ref{form}) is  computationally easier to handle than an 
ansatz of the type $\prod_{i<j}f(|x_{i}-x_{j}|)$, which might appear 
more natural in view of the heuristic remarks at the end of the last 
subection.

The function $f$ is chosen to be
\begin{equation}\label{deff}
f(r)=\begin{cases}
f_0(r)/f_{0}(b)&\text{for $0\leq r\leq b$},\\
1&\textrm{for $r>b$},
\end{cases}
\end{equation}
with $f_{0}(r)=u_{0}(r)/r$. The estimates (\ref{upperbound}) and 
(\ref{upperbound2}) are 
obtained by somewhat lengthy computations similar as in 
\cite{dyson}, but making use of 
(\ref{partint}). For details we refer to \cite{LSY1999} and \cite{S1999}.

A test wave function with Dirichlet boundary condition may be obtained 
by localizing the wave function (\ref{wave}) on the length scale $L$. 
The energy cost per particle for this is ${\rm (const.)}/L^2$.
\hfill$\Box$	

\subsection{LOWER BOUND}
%%%%%%%%%%%%%%%%%%%%%%
In the beginning it was explained  why the lower bound
for the bosonic ground state energy of (\ref{ham}) is not easy to  obtain. 
The three different length scales for bosons will play a role in the 
proof below. 
\begin{itemize}
\item The scattering length $a$.
\item The mean particle distance $\rho^{-1/3}$.
\item The `uncertainty principle length' $\ell_{c}$, defined by
$\mu\ell_{c}^{-2}=e_{0}(\rho)$, i.e., $\ell_{c}\sim (\rho a)^{-1/2}$.
\end{itemize}

Our lower bound for $e_{0}(\rho)$ is as follows.
\begin{theorem}[\textbf{Lower bound in the thermodynamic limit}]\label{lbth}  
For a  positive potential $v$ with finite range and $Y$ small enough
\begin{equation}\label{lowerbound}\frac{e_{0}(\rho)}{4\pi\mu\rho a}\geq 
(1-C\, 
Y^{1/17})
\end{equation}
with $C$ a constant. If $v$ does not have finite range, but decreases at 
least as fast as 
$1/r^{3+\varepsilon}$ at infinity with some $\varepsilon>0$, then an analogous 
bound to (\ref{lowerbound})
holds, but with $C$ replaced by another constant and 1/17 by another 
exponent, both of which may depend on $\varepsilon$.
\end{theorem}
It should be noted right away that the error term $-C\, Y^{1/17}$ in
(\ref{lowerbound}) is of no fundamental significance and is
not believed to reflect the true state of affairs. Presumably, it 
does not even have the right sign. We mention in passing that
$C$ can be taken to be
$8.9$ \cite{S1999}.

As mentioned in the Introduction a lower bound on $E_{0}(N,L)$ for 
finite $N$ and $L$ is of importance for applications to inhomogeneous 
gases, and in fact we derive (\ref{lowerbound}) from such a bound. We 
state it in the following way:
\begin{theorem}[\textbf{Lower bound in a finite box}] \label{lbthm2} 
	For a  positive potential $v$ with finite range there is 
a $\delta>0$ such that the the ground state energy of (\ref{ham}) with Neumann 
conditions satisfies
\begin{equation}\label{lowerbound2}E_{0}(N,L)/N\geq 4\pi\mu\rho 
a \left(1-C\, 
Y^{1/17}\right)
\end{equation} 
for all $N$ and $L$ with $Y<\delta$ and $L/a>C'Y^{-6/17}$. Here 
$C$ and $C'$ are constants,
independent of $N$ and $L$. (Note that the condition on $L/a$
 requires in particular that $N$ must be large enough, 
 $N>\hbox{\rm (const.)}Y^{-1/17}$.) 
 As in Theorem \ref{lbth} such a bound, but possibly with other 
 constants and another 
 exponent for $Y$, holds also for potentials $v$ of infinite range
 decreasing faster than $1/r^3$ at infinity.
\end{theorem}

The first step in the proof of Theorem \ref{lbthm2} is a generalization of 
a lemma of Dyson, which allows us to replace $v$ by a `soft' potential, 
at the cost of sacrificing kinetic energy and increasing the 
effective range.

\begin{lemma}\label{dysonl} Let $v(r)\geq 0$ with finite range $R_{0}$. Let 
$U(r)\geq 0$ 
be any function satisfying $\int U(r)r^2dr\leq 1$ and $U(r)=0$ for $r<R_{0}$. 
Let 
${\mathcal B}\subset \R^3$ be star shaped with respect to $0$ (e.g.\ 
convex with $0\in{\mathcal B}$). Then for all differentiable 
functions $\psi$
\begin{equation}\label{dysonlemma}
	\int_{\mathcal B}\left[\mu|\nabla\psi|^2+\mfr1/2 
v|\psi|^2\right]
\geq \mu a \int_{\mathcal B} U|\psi|^2.\end{equation}
\end{lemma}
{\it Proof.}  Actually, (\ref{dysonlemma}) holds with $\mu |\nabla \phi 
(\x)|^2$
replaced by the (smaller) radial kinetic energy,
 $\mu |\partial \phi (\x)/ \partial r|^2$, and  it  suffices to 
prove
the analog of (\ref{dysonlemma}) for the integral along each radial
line with fixed angular variables. Along such a line we write 
$\phi(\x) = u(r)/r$ with $u(0)=0$. We consider first the special case 
when when $U$ is a delta-function at some radius $R\geq 
R_0$,
i.e., \begin{equation}\label{deltaU}U(r)=\frac{1}{ 
R^2}\delta(r-R).\end{equation}
For such $U$ the analog of (\ref{dysonlemma}) along the radial line is
\begin{equation}\label{radial}\int_{0}^{R_{1}}
	\{\mu[u'(r)-(u(r)/r)]^2+\mfr1/2v(r)|u(r)]^2\}dr\geq
	\begin{cases}
		0&\text{if $R_{1}<R$}\\
			\mu a|u(R)|^2/R^2&\text{if $R\leq R_{1}$}
\end{cases}
\end{equation}
where $R_{1}$ is the length of the radial line segment in ${\mathcal 
B}$.
The case $R_{1}<R$ is trivial, 
because $\mu|\partial \psi/\partial r|^2+\mfr1/2 v|\psi|^2\geq 0$. 
(Note that positivity of $v$ is used here.) If $R\leq R_{1}$ we 
consider the integral on the the left side of (\ref{radial}) from 0 to $R$ 
instead of $R_{1}$ and
minimize it under the boundary condition that $u(0)=0$ 
and $u(R)$ is a fixed constant. Since everything is homogeneous in $u$ we may 
normalize this value to $u(R)=R-a$. 
This minimization problem leads to the zero energy 
scattering 
equation (\ref{scatteq}). Since $v$ is positive, the
solution is a true minimum and not just a 
stationary point.

Because $v(r)=0$ for $r>R_{0}$ the solution, $u_{0}$, satisfies $u_{0}(r)=r-a$ 
for $r>R_{0}$.  By partial integration, 
\begin{equation}\int_{0}^{R}\{\mu[u'_{0}(r)-(u_{0}(r)/r)]^2+
	\mfr1/2v(r)|u_{0}(r)]^2\}dr=\mu a|R-a|/R\geq \mu
a|R-a|^2/R^2.
	\end
{equation}
But $|R-a|^2/R^2$ is precisely 
the right side of (\ref{radial}) if $u$ satisfies the normalization condition.

This derivation of (\ref{dysonlemma}) for the special case (\ref{deltaU}) 
implies the 
general case, because every $U$ can be written as a 
superposition of  $\delta$-functions, 
$U(r)=\int R^{-2}\delta(r-R)\,U(R)R^2 dR$, and $\int U(R)R^2 dR\leq 1$ 
by assumption. 
\hfill$\Box$

By dividing $\Lambda$ for given points $\x_{1},\dots,\x_{N}$ into Voronoi 
cells  ${\mathcal B}_{i}$ that contain all points 
closer to $\x_{i}$ than to $\x_{j}$ with $j\neq i$ (these 
cells are star shaped w.r.t. $\x_{i}$, indeed convex), the 
following corollary of Lemma \ref{dysonl} can be derived in the same 
way as the corresponding  Eq.\ (28) in \cite{dyson}.

\begin{corollary} For any $U$ as in Lemma \ref{dysonl}
\begin{equation}\label{corollary}H_{N}\geq \mu a W\end{equation}
with
\begin{equation}\label{W}W(\x_{1},\dots,\x_{N})=\sum_{i=1}^{N}U(t_{i}),
\end{equation}
where $t_{i}$ is the distance of $\x_{i}$ to its {\it nearest 
neighbor} among the other points $\x_{j}$, $j=1,\dots, N$, i.e.,
\begin{equation}t_{i}(\x_{1},\dots,\x_{N})=\min_{j,\,j\neq 
i}|\x_{i}-\x_{j}|.\end{equation}
\end{corollary}
\noindent
(Note that $t_{i}$ has here a slightly different meaning than in 
(\ref{form}), where it denoted the distance to the nearest neighbor 
among the $\x_{j}$ with $j\leq i-1$.)

Dyson considers in \cite{dyson} a one parameter family of $U$'s that 
is essentially the same as the following choice, which is convenient for the 
present purpose:
\begin{equation}U_{R}(r)=\begin{cases}3(R^3-R_{0}^3)^{-1}&\text{for 
$R_{0}<r<R$ }\\
0&\text{otherwise.}
\end{cases}
\end{equation} 
We denote the corresponding interaction (\ref{W}) by $W_R$. For the hard core 
gas one obtains
\begin{equation}\label{infimum}E(N,L)\geq \sup_{R}\inf_{(\x_{1},\dots,\x_{N})} 
\mu a
W_R(\x_{1},\dots,\x_{N})\end{equation}
where the infimum is over $(\x_{1},\dots,x_{N})\in\Lambda^{N}$ with 
$|\x_{i}-\x_{j}|\geq R_{0}=a$, 
because of the hard core. At fixed $R$ simple geometry gives
\begin{equation}\label{fixedR}\inf_{(\x_{1},\dots,\x_{N})}
W_R(\x_{1},\dots,\x_{N})\geq \left(\frac{A}{R^3}-\frac{B}{ \rho 
R^6}\right)\end{equation}
with certain constants $A$ and $B$. An evaluation of these constants 
gives Dyson's bound
\begin{equation}E(N,L)/N\geq \frac{1}{10\sqrt 2} 4\pi\mu \rho 
a.\end{equation}

The main reason this method does not give a better bound is that $R$ 
must be chosen quite big, namely of the order of the mean particle 
distance $\rho^{-1/3}$, in order to guarantee 
that the spheres 
of radius $R$ around the $N$ points overlap. Otherwise the infimum of 
$W_R$ will be zero. But large $R$ means that $W_R$ is 
small. It should also be noted that this method does not work 
for potentials other than hard spheres: If $|\x_{i}-\x_{j}|$ 
is allowed to be less than $R_{0}$, then the right side of 
(\ref{infimum}) is zero because $U(r)=0$ for $r<R_{0}$.

For these reasons we take another route.
We still use  Lemma \ref{dysonlemma} to get into the soft potential regime, 
but we do {\it  not} 
sacrifice {\it all} the 
kinetic energy as in (\ref{corollary}). Instead we
write, for $\varepsilon>0$
\begin{equation}
	H_{N}=\varepsilon H_{N}+(1-\varepsilon)H_{N}\geq \varepsilon 
T_{N}+(1-\varepsilon)H_{N}
\end{equation}
with $T_{N}=-\sum_{i}\Delta_{i}$ and use (\ref{corollary}) only for the
part $(1-\varepsilon)H_{N}$. This gives
\begin{equation}\label{halfway}H_{N}\geq \varepsilon T_{N}+(1-\varepsilon)\mu 
a
W_R.\end{equation}
We consider the operator on the right side 
from the viewpoint of first order perturbation theory, 
with $\varepsilon T_{N}$ as the unperturbed  part, denoted $H_{0}$.

The ground state of $H_{0}$ in a box of side length $L$ is
$\Psi_{0}(\x_{1},\dots,\x_{N})\equiv L^{-3N/2}$ and we denote 
expectation values in this state by $\langle\cdot\rangle_{0}$.
A  computation, cf.\ Eq.\ (21) in \cite{LY1998}, gives
\begin{eqnarray}\label{firstorder}4\pi\rho\left(1-\mfr1/N\right)&\geq&
\langle W_R\rangle_{0}/N  \nonumber   \\ &\geq& 4\pi\rho
\left(1-\mfr1/N\right)\left(1-\mfr{2R}/L\right)^3
\left(1+4\pi\rho(1-\mfr1/N)(R^3-R_{0}^3)/3)\right)^{-1}.
\end{eqnarray}
The rationale behind the various factors is as follows: $(1-\mfr1/N)$ comes 
from 
the fact that the number of pairs is $N(N-1)/2$ and not $N^2/2$, 
$(1-{2R}/L)^3$ 
takes into account the fact that the particles do not interact beyond the 
boundary of 
$\Lambda$, and the last factor measures the probability to find another 
particle 
within the interaction range of the potential $U_R$ for a given particle.

The first order result (\ref{firstorder}) looks  at first sight quite  
promising, for if we let 
$L\to \infty$, $N\to \infty$ with $\rho=N/L^3$ fixed, and 
subsequently take
$R\to\infty$, then  $\langle W_R\rangle_{0}/N$ converges to $4\pi\rho$, which 
is 
just what is desired.
But the first order result (\ref{firstorder}) is not a 
rigorous bound on $E_0(N,L)$, we need
{\it error estimates}, and these will depend on $\varepsilon$, $R$ 
and $L$.

We now recall {\it Temple's inequality} \cite{TE} for the expectations 
values of an operator $H=H_{0}+V$ in the ground state 
$\langle\cdot\rangle_{0}$ of $H_{0}$. It is a simple 
consequence of the operator inequality
\begin{equation}(H-E_{0})(H-E_{1})\geq 0\end{equation}
for the two lowest eigenvalues, $E_{0}<E_{1}$, of 
$H$ and reads
\begin{equation}\label{temple}E_{0}\geq \langle H\rangle_{0}-\frac{\langle 
H^2\rangle_{0}-\langle 
H\rangle_{0}^2}{E_{1}-\langle H\rangle_{0}}\end{equation}
provided $E_{1}-\langle H\rangle_{0}>0$.
Furthermore, if $V\geq 0$ we may use $E_{1}\geq E_{1}^{(0)}$= second lowest 
eigenvalue of $H_{0}$ and replace $E_{1}$ in (\ref{temple}) by $E_{1}^{(0)}$.

%BEGIN CHANGE
{}From (\ref{firstorder}) and (\ref{temple}) we get the estimate
\begin{equation}\label{estimate2}\frac{E_{0}(N,L)}{ N}\geq 4\pi \mu a\rho
\left(1-{\mathcal 
E}(\rho,L,R,\varepsilon)\right)\end{equation}
with
\begin{eqnarray}\label{error}1-{\mathcal 
E}(\rho,L,R,\varepsilon)&=&(1-\varepsilon)\left(1-\mfr1/{\rho 
L^3}\right)\left(1-\mfr{2R}/L\right)^3
\left(1+\mfr{4\pi}/3\rho(1-\mfr1/N)(R^3-R_{0}^3))\right)^{-1}\nonumber\\ 
&\times&\left(1-\frac{\mu a\big(\langle 
W_R^2\rangle_0-\langle W_R\rangle_0^2\big)}{\langle 
W_R\rangle_0\big(E_{1}^{(0)}-\mu a\langle W_R\rangle_0\big)}\right).
\end{eqnarray}
%END CHANGE
To evaluate this further one may use the estimates (\ref{firstorder}) and the 
bound
\begin{equation}\label{square}
\langle W_R^2\rangle_0\leq 3\frac N{R^3-R_0^3}\langle W_R\rangle_0  
\end{equation}
which follows from $U_R^2=3({R^3-R_0^3})^{-1}U_R$ together with the 
Cauchy-Schwarz 
inequality. A glance at the form of the error term reveals, however, that it 
is 
{\it not} possible here to take the thermodynamic limit $L\to\infty$ with 
$\rho$ 
fixed:
We have $E_{1}^{(0)}=\varepsilon\pi\mu/L^2$ (this is the kinetic energy of a 
{\it single} particle in the first excited state in the box), and the factor
$E_{1}^{(0)}-\mu a\langle W_R\rangle_0$ in the denominator in (\ref{error}) 
is, 
up to unimportant constants and lower order terms, $\sim (\varepsilon 
L^{-2}-a\rho^2L^3)$. Hence the denominator eventually becomes negative and 
Temple's inequality looses its validity if $L$ is large enough.

As a way out of this dilemma we divide the big box $\Lambda$ into cubic {\it
cells} of side length $\ell$ that is kept {\it fixed} as $L\to \infty$.  The
number of cells, $L^3/\ell^3$, on the other hand, increases with $L$.  The $N$
particles are distributed among these cells, and we use (\ref{error}), with 
$L$
replaced by $\ell$, $N$ by the particle number, $n$, in a cell and $\rho$ by
$n/\ell^3$, to estimate the energy in each cell with {\it Neumann} conditions 
on the boundary.  This boundary condition leads to lower energy than any other
boundary condition.  For each distribution of the particles we add the
contributions from the cells, neglecting interactions across boundaries.  
Since
$v\geq 0$ by assumption, this can only lower the energy.  Finally, we minimize
over all possible choices of the particle numbers for the various cells 
adding up to $N$.  The energy obtained in this way is a lower bound to 
$E_0(N,L)$,
because we are effectively allowing discontinuous test functions for the
quadratic form given by $H_N$.

In mathematical terms, the cell method leads to 
\begin{equation}\label{sum}
E_0(N,L)/N\geq(\rho\ell^3)^{-1}\inf \sum_{n\geq 0}c_nE_0(n,\ell)
\end{equation}
where the infimum is over all choices of coefficients $c_n\geq 0$ (relative 
number of cells containing exactly $n$ particles), satisfying the constraints
\begin{equation}\label{constraints}
\sum_{n\geq 0}c_n=1,\qquad \sum_{n\geq 0}c_n n=\rho\ell^3.
\end{equation}

The minimization problem for the distributions of the particles among the 
cells would be easy if we knew that the ground state energy $E_0(n,\ell)$ (or 
a 
good
lower bound to it) were convex in $n$.  Then we could immediately conclude 
that
it is best to have the particles as evenly distributed among the boxes as
possible, i.e., $c_n$ would be zero except for the $n$ equal to the 
integer closest to 
$\rho\ell^3$. This would give 
\begin{equation}\label{estimate3}\frac{E_{0}(N,L)}{ N}\geq 4\pi \mu a\rho
\left(1-{\mathcal E}(\rho,\ell,R,\varepsilon)\right)\end{equation} i.e.,
replacement of $L$ in (\ref{estimate2}) by $\ell$, which is independent of 
$L$.
The blow up of ${\mathcal E}$ for $L\to\infty$ would thus be avoided.

Since convexity of $E_0(n,\ell)$ is not known (except in the thermodynamic 
limit) 
we must resort to other means to show that $n=O(\rho\ell^3)$ in all 
boxes. The rescue 
comes from {\it superadditivity} of $E_{0}(n,\ell)$, i.e., the property
\begin{equation}\label{superadd}
 E_0(n+n',\ell)\geq E_0(n,\ell)+E_0(n',\ell)
\end{equation}
which follows immediately from $v\geq 0$ by dropping the interactions between 
the $n$ particles and the $n'$ particles. The bound (\ref{superadd}) implies 
in 
particular that for any $n,p\in{\mathbb N}$ with $n\geq p$
\begin{equation}\label{superadd1}
E(n,\ell)\geq [n/p]\,E(p,\ell)\geq \frac n{2p}E(p,\ell)
\end{equation}
since the largest integer $[n/p]$ smaller than $n/p$ is in any case $\geq 
n/(2p)$.

The way (\ref{superadd1}) is used is as follows:
Replacing $L$ by $\ell$, $N$ by $n$ and $\rho$ by $n/\ell^3$ in 
(\ref{estimate2})  we have for fixed $R$ and $\varepsilon$
\begin{equation}\label{estimate4}
E_{0}(n,\ell)\geq\frac{ 4\pi \mu a}{\ell^3}n(n-1)K(n,\ell)
\end{equation}
with a certain function $K(n,\ell)$ determined by (\ref{error}). We 
shall see that $K$ is monotonously decreasing in $n$, so that if 
$p\in{\mathbb N}$  and $n\leq p$ then
\begin{equation}\label{n<p}
E_{0}(n,\ell)\geq\frac{ 4\pi \mu a}{\ell^3}n(n-1)K(p,\ell).
\end{equation}
We now split the sum in (\ref{sum}) into two parts. 
For $n<p$ we use (\ref{n<p}), and for $n\geq p$ we use (\ref{superadd1}) 
together with (\ref{n<p}) for $n=p$. The task is thus to minimize
\begin{equation}\label{task}
\sum_{n<p}c_n n(n-1)+\mfr1/2\sum_{n\geq p}c_nn(p-1)
\end{equation}
subject to the constraints ({\ref{constraints}). 
Putting 
\begin{equation}
k:=\rho\ell^3 \quad\text{and}\quad t:=\sum_{n<p}c_n n\leq k
\end{equation}
we have $\sum_{n\geq p}c_n n=k-t$, and since 
$n(n-1)$ is convex in $n$, and $\sum_{n<p}c_n\leq 1$ the expression 
(\ref{task})
is
\begin{equation}
\geq t(t-1)+\mfr1/2(k-t)(p-1).
\end{equation}
We have to minimize this for $1\leq t\leq k$. If $p\geq 4k$ the minimum is 
taken 
at $t=k$ and is equal to $k(k-1)$. Altogether we have thus shown that
%BEGIN CHANGE
\begin{equation}\label{estimate1}
\frac{E_{0}(N,L)}{ N}\geq 4\pi \mu a\rho\left(1-\frac1{\rho\ell^3} \right) 
K(4\rho\ell^3,\ell).
\end{equation}
%END CHANGE

What remains is to take a closer look at $K(4\rho\ell^3,\ell)$, which depends 
on 
the parameters $\varepsilon$ and $R$ besides $\ell$, and choose the parameters 
in an optimal way. 
%BEGIN CHANGE
>From (\ref{error}) and 
(\ref{square}) we obtain
\begin{eqnarray}\label{Kformula}
K(n,\ell)&=&(1-\varepsilon) \left(1-\mfr{2R}/\ell\right)^3
\left(1+\mfr{4\pi}/3\rho(1-\mfr1/n)(R^3-R_{0}^3))\right)^{-1}
\nonumber
\\ &\times&\left(1-\frac3\pi
\frac{an}{(R^3-R_{0}^3)(\varepsilon\ell^{-2}-4a\ell^{-3}n(n-1))}\right).	
\end{eqnarray}
The estimate (\ref{estimate4}) with this $K$ is valid as long as the 
denominator in the last factor
%END CHANGE
in (\ref{Kformula}) is $\geq 0$, and in order to have a formula 
for 
all $n$ we can take 0 as a 
trivial lower bound in other cases or when (\ref{estimate4}) is 
negative. As required
for (\ref{n<p}), $K$ is monotonously decreasing in $n$. We now insert
$n=4\rho\ell^3$ and obtain
%BEGIN CHANGE
\begin{eqnarray}\label{Kformula2}
K(4\rho\ell^3,\ell)&\geq&(1-\varepsilon)\left(1-\mfr{2R}/\ell\right)^3
\left(1+({\rm const.})Y(\ell/a)^3 (R^3-R_{0}^3)/\ell^3\right)^{-1}
\nonumber
\\ &\times&\left(1-
\frac{\ell^3}{(R^3-R_{0}^3)}\frac{({\rm const.})Y}
{(\varepsilon(a/\ell)^{2}-({\rm const.})Y^2(\ell/a)^3)}\right)	
\end{eqnarray}
with $Y=4\pi\rho a^3/3$ as before. Also, the factor 
\begin{equation}
\left(1-\frac1{\rho\ell^3} \right)=(1-({\rm const.})Y^{-1}(a/\ell)^{3})
\end{equation}
%END CHANGE
in (\ref{estimate1})
(which is the ratio between
$n(n-1)$ and $n^2$) must not be be forgotten. We now make the ansatz
\begin{equation}\label{ans}
\varepsilon\sim Y^\alpha,\quad a/\ell\sim Y^{\beta},\quad 
(R^3-R_{0}^3)/\ell^3\sim Y^{\gamma}	
\end{equation}	
with exponents $\alpha$, $\beta$ and $\gamma$ that we choose 
in an optimal way. The conditions to be met are as follows:
%BEGIN CHANGE
\begin{itemize}
\item $\varepsilon(a/\ell)^{2}-({\rm const.})Y^2(\ell/a)^3>0$. This 
holds for all small enough $Y$, provided 
$\alpha+5\beta<2$ which follows from the conditions below.
\item $\alpha>0$ in order that $\varepsilon\to 0$ for $Y\to 0$.
\item $3\beta-1>0$ in order that  $Y^{-1}(a/\ell)^{3}\to 0$ for for $Y\to 
0$. 
\item $1-3\beta+\gamma>0$ in order that  
$Y(\ell/a)^{3}(R^3-R_{0}^3)/\ell^3\to 0$ for for $Y\to 0$.
%END CHANGE
\item $1-\alpha-2\beta-\gamma>0$ to control the last factor in 
(\ref{Kformula2}).
\end{itemize}
Taking
\begin{equation}\label{exponents}
\alpha=1/17,\quad \beta=6/17,\quad \gamma=3/17	
\end{equation}
all these conditions are satisfied, and
%BEGIN CHANGE
\begin{equation}
\alpha=	3\beta-1=1-3\beta+\gamma=1-\alpha-2\beta-\gamma=1/17.
\end{equation}
It is also clear that 
$2R/\ell\sim Y^{\gamma/3}=Y^{1/17}$, up to higher order terms.
%END CHANGE
This completes the proof of Theorems 3.1 and 3.2, for the case 
of potentials with  finite range. By optimizing the proportionality 
constants in (\ref{ans}) one can show that $C=8.9$ is possible in Theorem 1.1 
\cite{S1999}. The extension to potentials of infinite range 
decreasing faster than $1/r^3$ at infinity is 
obtained by approximation by finite range potentials, controlling the 
change of the scattering length as the cut-off is removed. See 
Appendix B in \cite{LSY1999} for details. A slower decrease than 
$1/r^3$ implies infinite scattering length. \hfill$\Box$

The exponents (\ref{exponents}) mean in particular that
\begin{equation}a\ll R\ll \rho^{-1/3}\ll \ell \ll(\rho 
a)^{-1/2},\end{equation}
whereas Dyson's method required $R\sim \rho^{-1/3}$ as already explained. 
The condition $\rho^{-1/3}\ll \ell$ is required in order to have many 
particles in each box and thus $n(n-1)\approx n^2$. The condition
$\ell \ll(\rho a)^{-1/2}$ is necessary for a spectral gap
gap $\gg e_{0}(\rho)$ in Temple's inequality. It is also clear that 
this choice of $\ell$  would lead to a far too big
energy and no bound for $ e_{0}(\rho)$ if we had chosen Dirichlet instead of 
Neumann boundary 
conditions for the cells. But with the latter the method works!
%%%%%%%%%%

\bigskip

\section{THE DILUTE BOSE GAS IN 2D} \label{sect2d}

The two-dimensional theory, in contrast, began to receive attention
only relatively late. The first derivation of the  asymptotic formula
was, to our knowledge, done by Schick \cite{schick}, as late as 1971!
He found

\begin{equation}
    e_0(\rho)  \simeq \frac{4\pi \mu \rho}{ |\ln(\rho a^2) |}.
\label{2den}
\end{equation}

The scattering length $a$ in  (\ref{2den}) is defined using the zero
energy scattering equation (\ref{3dscatteq}) but instead of
$\psi(r)\approx 1-a/r$ we now impose the asymptotic condition
$\psi(r)\approx \ln(r/a)$. 
This is explained  in the appendix to \cite{LY2d}. 

Note that the answer could not possibly be $e_0(\rho)  \simeq 4\pi \mu 
\rho a $ because that would be dimensionally wrong. But $e_0(\rho) $ must
essentially be proportional to $\rho$, which leaves no room for an $a$
dependence --- which is ridiculous! It turns out that this dependence
comes about in the $\ln(\rho a^2)$ factor.

One of the intriguing facts about (\ref{2den}) is that the energy for $N$
particles is {\it not equal} to $N(N-1)/2$  times the energy for two
particles in the low density limit --- as is the case in
three dimensions.  The latter quantity,  $E_0(2,L)$, is, asymptotically,
for large $L$, equal to $8\pi \mu L^{-2} \left[ \ln(L^2/a^2)
\right]^{-1}$.  Thus, if the  $N(N-1)/2$ rule were to apply in 2D, (\ref{2den})
would have to be replaced by the much smaller quantity $4\pi \mu
\rho\left[ \ln(L^2/a^2) \right]^{-1}$. In other words, $L$, which tends
to $\infty$ in the thermodynamic limit, has to be replaced by the mean
particle separation, $\rho^{-1/2}$ in the logarithmic factor. Various
poetic formulations of this curious fact have been given, but it 
remains true that the non-linearity is something that does not  occur in more
than two-dimensions and its precise nature is hardly obvious,
physically. This anomaly is the main reason that the 2D
result is not a trivial extension of \cite{LY1998}.

The proof of (\ref{2den}) is in \cite{LY2d}.
The (relative) error terms to (\ref{2den}) givein in 
\cite{LY2d} are $|\ln(\rho a^2) |^{-1}$
for the upper bound and     $|\ln(\rho a^2) |^{-1/5}$

To prove (\ref{2den}) the essential new step is to modify Dyson's lemma
for 2D. The rest of the proof parallels that for 3D. The 2D version of
Lemma \ref{dysonl} is \cite{LY2d}:

\begin{lemma}\label{dyson2d} Let $v(r)\geq 0$ with finite
range $R_{0}$. Let $U(r)\geq 0$ be any function satisfying $\int
U(r)\ln(r/a)rdr\leq 1$ and $U(r)=0$ for $r<R_{0}$.  Let ${\mathcal
B}\subset \R^3$ be star shaped with respect to $0$ (e.g.\ convex with
$0\in{\mathcal B}$). Then for all differentiable functions $\psi$
\begin{equation}\label{dysonlem2d}
        \int_{\mathcal B}\left[\mu|\nabla\psi|^2+\mfr1/2
v|\psi|^2\right] \geq \mu  \int_{\mathcal B} U|\psi|^2.\end{equation}
\end{lemma}

\bigskip

\section{BOSE-EINSTEIN CONDENSATION}\label{sectbe}

Let us comment very briefly on the notion of Bose-Einstein condensation
(BEC). Given the normalized ground state wave function 
$\Psi_{0}(\x_{1},\dots,\x_{N})$ we can form the one-body density matrix
which is an operator
on $L^2(\R^n)$    ($n=2$ or $3$) given by the kernel
$$
\gamma(\x,\, \y) =N \int_{\textrm{BOX}^{N-1}} \Psi_{0}(\x,\, \x_{2},\dots,\x_{N})
\Psi_{0}(\y, \, \x_{2},\dots, \x_{N}) d\x_{2}\cdots d\x_{N}.
$$
%\bigskip
Then $\int \gamma(\x,\, \x) d\x =\textrm{Trace}(\gamma) = N$.
BEC is the assertion that this operator has an eigenvalue of order $N$.
Since $\gamma$ is  a positive kernel and, hopefully, translation invariant
in the thermodynamic limit, the eigenfunction belonging to  the largest
eigenvalue must be the constant function $(\textrm{volume})^{-1/2}$.
Therefore, another way to say that BEC exists is that
$$
\int\int \gamma(\x,\, \y) d\x d\y = \textrm{O}(N). 
$$

Unfortunately, this is something that is frequently invoked but never
proved --- except for one special case:  hard core bosons on a
lattice at half-filling (i.e., 
$N=$ half the number of lattice sites). The proof is in \cite{KLS}.

The problem remains open after about 70 years. It is not at all clear
that BEC is essential for superfluidity, as frequently claimed. 
Our construction in section \ref{sect3d} shows that BEC exists on
a length scale of order $\rho^{-1/3} Y^{ -1/17}$ which, unfortunately,
is not a `thermodynamic' length like $\textrm{volume}^{1/3}$.

\section{GROSS-PITAEVSKII EQUATION FOR TRAPPED BOSONS} \label{sectgp}

In the recent experiments on Bose condensation, the particles have to be 
confined in a cold `trap' instead of a `box' and we are
certainly not at the `thermodynamic limit'. 
For a `trap' we add a {\it slowly} varying confining potential
$ V$, with $V(\x)\to \infty $ as $|\x|\to \infty$ .
The Hamiltonian becomes
\begin{equation}\label{trapham}
H =  \sum_{i=1}^{N} -\mu \Delta_i +V(\x_i) +
 \sum_{1 \leq i < j \leq N} v(|\x_i - \x_j|)
\end{equation}
If $v=0$, then
$$\Psi_{0}(\x_{1},\dots,\x_{N})=\hbox{$\prod_{i=1}^{N}$}\Phi_{0}(\x_{i})
$$
with $\Phi_{0}$= normalized ground state of $- \Delta + V(\x)$
with eigenvalue $=\lambda$.

The idea is to use the information about the thermodynamic limiting
energy of the dilute Bose gas in a box to find the ground state 
energy of  (\ref{trapham}). This has been done
with Robert Seiringer and Jakob Yngvason \cite{LSY1999, LSY2d}. 
% A longer exposition of this joint work with 
%Robert Seiringer and Jakob Yngvason \cite{LSY1999, LSY2d} is given in the
%paper by Seiringer in this proceedings.

As we saw in Sections \ref{sect3d} and
\ref{sect2d} there is a difference in the $\rho$ dependence between
two and three dimensions, so we can expect a related difference now.
We discuss 3D first.

\subsection{THREE-DIMENSIONS}

Associated with the quantum mechanical ground state energy  problem 
is the GP energy functional
\begin{equation}\label{gpfunc3d}
\E^{\rm GP}[\Phi]=\int_{R^3}\left(\mu|\nabla\Phi|^2+V|\Phi|^2+4\pi
\mu a|\Phi|^4\right)d^3\x
\end{equation}
with the subsidiary condition
$$\int_{R^3}|\Phi|^2=N$$
and corresponding energy
\begin{equation}\label{gpen3d}
E^{\rm GP}(N,a)=\inf \{\E^{\rm GP}[\Phi] \ :\
{\int|\Phi|^2=N}\}=
\E^{\rm GP}[\Phi^{\rm GP}].
\end{equation}

As before, $a$ is the scattering length of $v$.

It is not hard to prove that for every choice of the real number $N$
there is a unique minimizer for $\Phi^{\rm GP}$ for $\E^{\rm GP} $.

\bigskip

\subsection*{\textsf{ Relation of $\E^{\rm GP}$ and $E_0$:}}
If $v=0$ then  clearly $\Phi^{\rm GP}=\sqrt{N}\ \Phi_{0}$, and then
$\E^{\rm GP}=N\lambda = E_0$.  In the other extreme, if $V(\x)=0$ for
$\x$ inside a large box of volume $L^3$ and $V(\x)= \infty$ otherwise,
then we take $\Phi^{\rm GP} \approx \sqrt{N/L^3}$ and we get $E^{\rm
GP}(N) = 4\pi \mu a N^2/L^3=$ previous, homogeneous $E_0$ in the low
density regime. (In this case, the gradient term in $\E^{\rm GP}[\Phi]$
plays no role.)

In general, we expect that $E^{\rm GP} = E_0$ in a suitable limit.
This limit has to be chosen so that {\it all three} terms
in $\E^{\rm GP}[\Phi]$ make a contribution. It turns  out that fixing
$Na $ is the right thing to do (and this can be quite good experimentally
since $Na$ can range from about 1 to 1000).

\begin{theorem} \label{thmgp3} If $a_{1}=Na$ is fixed,
$$\lim_{N\to\infty}\frac{{E_0(N,a)}}{ {E^{\rm
GP}(N,a)}}=1$$

Moreover, the GP density is a limit of the QM density 
 (convergence in weak $L_1$-sense):
$$\lim_{N\to\infty}\frac{1}{ N}\rho^{\rm QM}_{N,a}(\x)=
\left|{\Phi^{\rm GP}_{1,Na}}(\x)\right|^2\ .$$
\end{theorem} 

We could imagine, instead, an $N \to \infty$ limit in which $a\gg N^{-1}$,
i.e. $a_1\to\infty$, but still $\bar\rho a^3\to 0$,  where $\bar\rho$ is
the average density, given by
\begin{equation}\label{rhobar}
\bar\rho = \frac{1}{N}\int (\Phi^{\rm GP})^4 .
\end{equation}
In this case we  simply
omit the $\int|\nabla\Phi|^2$ term in $\mathcal{E}^{\rm GP}$; this theory
is usually called `Thomas-Fermi theory', but it has nothing to do with
the fermionic theory invented by Thomas and Fermi in 1927. It is
appropriate for the case in which $Na$ is much bigger than 1, e.g.,
$Na \approx 1000$.

\begin{proof} \textsf{Outline:} Getting an upper bound for $E_0$ in terms of
$E^{\rm GP}$ is relatively easy, as
before. The problem is the lower bound.

One might suppose that one decomposes $R^3$ into small boxes as before,
with Neumann b.c. on each box, and in each box one approximates $V(\x)$
by a constant.

This will NOT work, even if $v=0$, because all the
particles will then want to be in the box with the smallest value of
$V$. The gradient term will vanish and we will not get $E^{GP}=N\lambda$.

The trick is to write the quantum $\Psi_0$ as
$$
\Psi_{0}=\hbox{$\prod_{i=1}^N$}\Phi^{\rm GP}(\x_{i})F(\x_{1},
\dots,\x_{N}).
$$

This leads to a variational problem for $F$ instead of
$\Psi_{0}$.
Partial integration and the variational equation for $\Phi^{\rm GP}$
lead to the  replacements:
\noindent
$$\hbox{External potential: } V(\x)\quad \rightarrow \quad
 -8\pi a |\Phi^{\rm GP}(\x)|^2$$
and
\noindent
$$\hbox{Measure: }\hbox{$\prod_{i=1}^N$}d\x_i
\quad \rightarrow \quad \hbox{$\prod_{i=1}^N$}
|\Phi^{\rm GP}(\x_i)|^2 d\x_{i}.
$$
We have to show that, with these replacements, the energy is bounded below by
$-4\pi \mu a \int |\Phi^{\rm GP}|^4$ (up to small errors).
{\it Now} we can effectively
use  the Neumann box method on {\it this} $F$ problem.

There are still plenty of technical difficulties, but the back of the
problem has been broken by
recasting it in terms of $F$ instead of $\Psi_{0}$.
\end{proof}

\subsection{TWO-DIMENSIONS}

In view of the two-dimensional result (\ref{2den}) we might suppose that 
we must replace $4\pi \mu a \Phi^4$ in (\ref{gpfunc3d}) by 
$4 \pi  \mu \Phi^4 / |\ln (\Phi^2 a^2)|$, as suggested by 
(\ref{2den}).  This is indeed correct, and a limit  theorem for the
energy and density is obtained that is similar to Theorem \ref{thmgp3}.
There is a difference, however. This time one has to fix the ratio
$N/\ln a$ instead of $Na$. 

Since the logarithm is such a slowly varying function it turns
out that one can still get the correct limit and yet simplify the
problem by replacing $|\ln (\Phi^2 a^2)|^{-1}$ in the functional by
the constant
$|\ln (\bar\rho a^2)|^{-1}$, where $\bar\rho $ is the average density
in (\ref{rhobar}), but for 2D, of course.

\bigskip

\section{THE CHARGED BOSE GAS}

The setting now changes abruptly. Instead of particles interacting
with a short-range potential $v(\x_i-\x_j)$ they interact via the Coulomb
potential
$$v(|\x_i-\x_j|)  = |\x_i-\x_j|^{-1}
$$
(in 3 dimensions). There are $N$ particles in a large box of volume
$L^3$ as before, with $\rho =N/L^{3}$. We know from the work in
\cite{LN} that a nice thermodynamic limit exists for 
$e_0=E_0/N$.

To offset the huge Coulomb
repulsion (which would drive the particles to the walls of the box)
we add a uniform negative background of precisely the same charge,
namely density $\rho$. Our Hamiltonian is thus
\begin{equation}\label{foldyham}
H=  - \mu \sum_{i=1}^{N} \Delta_i -V(\x_i) +
 \sum_{1 \leq i < j \leq N} v(\x_i - \x_j)  +C
\end{equation}
with
$$
V(\x)=\rho \int_{\rm BOX} |\x-\y |^{-1}d^3y  \qquad \qquad
{\rm and}\qquad \qquad
 C= \frac{1}{ 2} \rho \int_{\rm BOX} V(\x)d^3 x\ .
$$

Despite the fact that the Coulomb potential is positive definite, each
particle interacts only with others and not with itself. Thus, $E_0$ can
be (and is) negative (just take $\Psi=$const). This time, \textit{
large} $\rho$ is the `weakly interacting' regime.

Another way in which this problem is different from the previous one
is that \textit{ perturbation theory is correct to leading order}. If
one computes $(\Psi, H \Psi)$ with  $\Psi=$const, one gets the right
first order answer, namely $0$. It is the next order in $1/\rho$ that
is interesting, and this is \textit{ entirely} due to correlations.
In 1961 Foldy \cite{FO} calculated this correlation energy according
to the prescription of Bogolubov's 1947 theory. That theory was not
exact for the dilute Bose gas, as we have seen, even to first order
. We are now looking at \textit{ second} order, which should be even
worse. Nevertheless, there was good physical intuition that this
calculation should be asymptotically \textit{ exact}. \textit{ It is!}
and this was recently proved with Jan Philip Solovej \cite{LS}.

The Bogolubov theory states that the main contribution to the
energy comes from pairing of particles into momenta $\k, -\k$ and is
the bosonic analogue of the BCS theory of superconductivity which came
a decade later. I.e., $\Psi_0$ is a sum of products of germs of the form
$\exp\{i\k \cdot (\x_i-\x_j)\}$.

Foldy's energy, based on Boglubov's ansatz, has now been proved. His
calculation yields an upper bound.  The lower bound is the hard part,
and Solovej and I do this using the decomposition into `Neumann boxes'
as above. But unlike the short range case, many complicated gymnastics
are needed to control the long range $|\x|^{-1} $ Coulomb potential.
The two bounds agree to leading order in $\rho$,
namely $\rho^{1/4}$ (although there is an unimportant
technical point that different boundary conditions are used for the two bounds).

\begin{theorem} For large $\rho$
\begin{equation}\label{foldyen}
e_0(\rho) \approx
-0.402 \, r_s^{-3/4} \frac{me^4}{\hbar^2}
\end{equation}
 where $r_s=(3/4\pi \rho)^{1/3}e^2m/\hbar^2$.
\end{theorem}

This is the {\it first example} (in more than 1 dimension) in which
Bogolubov's pairing theory has been rigorously validated. It has to be
emphasized, however, that Foldy and Bogolubov rely on the existence of
Bose-Einstein condensation. We neither make such a hypothesis nor does
our result for the energy imply the existence of such condensation. As
we said earlier, it is sufficient to prove condensation in small boxes
of fixed size.

Incidentally, the one-dimensional example for which Bogolubov's theory is
asymptotically exact to the first two orders  (high density) is the 
repulsive delta-function Bose gas \cite{LL} 

To appreciate the  $-\rho^{1/4}$ nature of (\ref{foldyen}), it is useful to
compare it with what one would get if the bosons had infinite mass,
i.e., the first term in (\ref{foldyham}) is dropped. Then the energy would
be proportional to $-\rho^{1/3}$ as shown in \cite{LN}. Thus, the effect
of quantum mechanics is to lower $\frac{1}{3}$ to $\frac{1}{4}$. 

It is supposedly true that there is a critical mass above which the 
ground state should show crystalline ordering (Wigner crystal), but this
has never been proved and it remains an intriguing open problem, even
for the infinite mass case. Since the relevant parameter is
$r_s$, large mass is the same as small $\rho$, and  is outside the
region where a Bogolubov approximation can be expected to hold.

Another important remark about the $-\rho^{1/4}$  law is its relation
to the $-N^{7/5}$ law for a $\mathit{two}$-component charged Bose gas.
Dyson \cite{D2} proved that the ground state energy for such a gas was at
least as negative as $-(\mathrm{const})N^{7/5}$  as $N\to \infty$.  Thus,
thermodynamic stability (i.e., a linear lower bound) fails for this gas.
Years later, a lower bound of this $-N^{7/5}$ form was finally established
in \cite{CLY}, thereby proving that this law is correct.  The connection
of this $-N^{7/5}$ law with the jellium $-\rho^{1/4}$ law (for which a
corresponding lower bound was also given in \cite{CLY}) was pointed out
by Dyson \cite{D2} in the following way. Assuming the correctness of
the $-\rho^{1/4}$  law, one can treat the 2-component gas by treating
each component as a background for the other. What should the density
be? If the gas has a radius $L$ and if it has $N$ bosons then $\rho =
N L^{-3}$. However, the extra kinetic energy needed to compress the
gas to this radius is $N L^{-2}$. The total energy is then $N L^{-2}
- N \rho^{1/4}$, and minimizing this with respect to $L$ leads to the
$-N^{7/5}$ law. A proof going in the other direction is in \cite{CLY}.

A problem somewhat related to bosonic jellium is \textit{fermionic} jellium. 
Graf and Solovej \cite{GS} have proved that the first two terms are what
one would expect, namely
$$
        e(\rho)=C_{TF}\rho^{5/3}-C_{D}\rho^{4/3}+o(\rho^{4/3}),
$$
where $C_{TF}$ is the usual Thomas-Fermi constant and $C_{D}$ is the
usual Dirac exchange constant. 

Let us conclude with a few more details about some of the
technicalities.

\subsection{FOLDY'S CALCULATION,  PAIRING THEORY AND IDEAS  IN THE
RIGOROUS PROOF} 
Foldy uses periodic boundary conditions for $-\Delta$, so the
problem is on a torus. In order to make the Coulomb potential periodic and
to take care of the background he
replaces $|x-y|^{-1}$ by
$$
        \sum_{p\ne0}L^{-3}|p|^{-2}\exp(ip(x-y)),
$$
where the sum is over `periodic momenta', $p$.
(Note the $p\ne0$ condition, so the spatial average of this potential is 0).

Foldy's Hamiltonian is
$$
        H'=\sum_{i=1}^N-{\textstyle\frac{1}{2}}\Delta_i
        +\sum_{i<j}\sum_{p\ne0}L^{-3}|p|^{-2}\exp(ip(x_i-x_j),
$$
in which the 
$p\ne0$ condition is supposed to make up for the  background that is 
not explicitly included in $H'$. It is `physically clear'
that this device works, but a rigorous proof of it is not easy.

The next step is to use the second quantization formalism, which is the
one used by Bogolubov and which is a very convenient bookkeeping device
(but it has to be noted that it is no more than a convenient device and
it does not introduce any new physics or mathematics). 
$$
 H'=\sum_p |p|^2a^*_pa_p
        +\sum_{p\ne0}L^{-3}|p|^{-2}\sum_{k,q}a^*_{k+p}a^*_{q-p}
        a_qa_k ,
$$
where the $ a_p$ operators satisfy  the usual bosonic commutation
relations. An important observation is that
since $p\ne 0$ there are no terms with 3 or 4 $a^\sharp_0$. This is
different from the situation with the usual short range potential
treated in Section \ref{sect3d}, and it means
that the leading term in perturbation theory vanishes. Everything now
comes from the terms with two $a^\sharp_0$.

\subsubsection{{The 
Bogolubov approximation:}} The 
motivation is Bose condensation: Almost all particles
are in the state of momentum $p=0$ created by
$a^*_0$. Thus:

\textsf{Step 1 in the Bogolubov approximation:} Keep only quartic
terms with precisely two
$a^\sharp_0$ (and ignore terms with one or no $a^\sharp_0$).
We obtain the reduced Hamiltonian
\begin{eqnarray*}
 H''&=&\sum_p |p|^2a^*_pa_p
        +\sum_{p\ne0}L^{-3}|p|^{-2}\Bigl[a^*_{p}a^*_{0}
        a_pa_0\\
        &&+a^*_{0}a^*_{-p}
        a_0a_{-p}+a^*_{p}a^*_{-p}a_0a_0+a^*_{0}a^*_{0}a_pa_{-p}\Bigr]
\end{eqnarray*}

\textsf{Step 2 in the Bogolubov approximation:} Replace the \textit{operators}
 $a^\sharp_0$ by the number
$\sqrt{N}$. We then obtain:
\begin{eqnarray*}
 H'''=&\displaystyle\sum_{p\ne0}& |p|^2a^*_pa_p
        +\rho|p|^{-2}\Bigl[a^*_{p}a_p+a^*_{-p}a_{-p}\\
        &&+a^*_{p}a^*_{-p}+a_pa_{-p}\Bigr]
\end{eqnarray*}
This is a quadratic Hamiltonian and can be diagonalized by completing
 the square:
\begin{eqnarray*}
        H'''&=&\sum_p A_p(a^*_p+\beta_pa_{-p})
        (a_p+\beta_pa_{-p}^*)\\
        &&+A_p(a^*_{-p}+\beta_pa_{p})
        (a_{-p}+\beta_pa_{p}^*)\\&&-2\sum_{p\ne0}A_p\beta_p^2 
\end{eqnarray*}
The last term comes from the commutator
$[a_p,a_q^*]=\delta_{pq}$.
\begin{eqnarray*}
        A_p(1+\beta_p^2)&=&\frac{1}{2}|p|^2+\rho|p|^{-2}\\
        2A_p\beta_p&=&\rho|p|^{-2}
\end{eqnarray*}
The ground state energy
is given by the last term above.
$$
 e=\lim_{L\to\infty}-\frac{2}{L^{3}}\sum_{p\ne0}A_p\beta_p^2
        =\int A_p\beta_p^2=C_F\rho^{5/4}.
$$
In this approximation the ground state wave function $\psi$ satisfies
$$(a_p+\beta_pa_{-p}^*)\psi=0,$$
for all $p\ne0$.

In the original language (in which $a_0$ an operator)
this corresponds to a function of the form
\begin{eqnarray*}
 \psi&=&1+\sum_{i<j}f(x_i-x_j)\\
        &&+c\sum_{\genfrac{}{}{0pt}{0}{i,j,l,k}
          {\mathrm{different}}}  \!\!\!\!  f(x_i-x_j)f(x_l-x_k)
        +\ldots
\end{eqnarray*}
where $\hat{f}(p)=\beta_p$.
In fact, $\hat{f}(p)=G(|p|^4/\rho)$, with $G$ independent
of $\rho$.
Thus $f$ varies on a length scale $\rho^{-1/4}$
(which is the typical inter\textit{pair} distance).

\subsubsection{{Ideas in the rigorous proof:}} As in the short
range case in Section \ref{sect3d}, there is no need to prove Bose
condensation globally. It is enough to do so on a short scale.

\begin{itemize}
\item Localize particles, by means of Neumann bracketing, in ``small''
boxes of size $\ell$. The constant function (i.e., the `condensate')
is not affected by this localization since the constant function in a 
small box satisfies Neumann boundary conditions. The function 
$f$ discussed above is not affected very much
if $\ell\gg \rho^{-1/4}$. We choose $\ell$ close to
$\rho^{-1/4}$.
\item We control the Coulomb interaction between boxes by
using an averaging method in \cite {CLY}. The error in neglecting
intercell interactions 
can be shown to be dominated by $N/\ell\ll N\rho^{1/4}$.
\item We establish condensation on the scale $\ell$ by noting that the
first non-zero Neumann eigenvalue is $\sim\ell^{-2}$.
If $N_+$  is the expected number $N_+$ of particles not in the 
condensate in the ``small box'', their energy is bounded from below by
$N_+ \ell^{-2}
\sim N_+ \rho^{1/2}$. If this is to be 
consistent with the total
energy $-N\rho^{1/4}$ we should expect to have
$N_+\ll N \rho^{-1/4}$, i.e., local condensation exists.
\item This is established by means of  a bootstrap procedure.
\item Having established local condensation one starts the hard work of
proving that the Bogolubov `quadratic' approximation really gives 
the leading term in the energy. There is however a new difficulty that
arises from the finite size of the small boxes in which we are working.
Neumann, and not periodic boundary conditions must  be used. Since we no
longer have `momentum conservation' in the small boxes, diagonalization
of the Hamiltonian by `completing
the square' is not the simple algebraic problem it was before. 
\item Nevertheless, it can all be carried to a succesful conclusion

\end{itemize}

\bibliographystyle{amsalpha}

\begin{thebibliography}{A}

%\begin{thebibliography}{99}

\bibitem{LY1998}
E.H. Lieb, J. Yngvason, \textit{ Ground State Energy of the low density
Bose Gas}, Phys. Rev. Lett. \textbf{80}, 2504--2507 (1998). arXiv
math-ph/9712138, mp\_arc 97-631

\bibitem{LY2d} E.H. Lieb, J. Yngvason, \textit{The Ground State Energy
of a Dilute Two-dimensional Bose Gas}, J. Stat. Phys. (in press) arXiv
math-ph/0002014, mp\_arc 00-63.

\bibitem{LSY1999} E.H. Lieb, R. Seiringer, and J. Yngvason, \textit{
Bosons in a Trap: A Rigorous Derivation of the Gross-Pitaevskii
Energy Functional},  Phys.  Rev A \textbf{ 61}, 043602-1 -- 043602-13
(2000)mp\_arc 99-312, arXiv math-ph/9908027 (1999).

\bibitem{LSY2d} E.H. Lieb, R. Seiringer and J. Yngvason, \textit{
A Rigorous Derivation of the Gross-Pitaevskii Energy Functional for a
Two-dimensional Bose Gas}.  arXiv cond-mat/0005026, mp\_arc 00-203.

\bibitem{LS} E.H. Lieb, J.P. Solovej, \textit{ Ground State Energy
of the One-Component Charged Bose Gas}. arXiv cond-mat/0007425, mp\_arc
00-303.

\bibitem{LYbham} E.H. Lieb, J. Yngvason, \textit{ The
Ground State Energy of a Dilute Bose Gas}, in \textit{ Differential
Equations and Mathematical Physics, University of Alabama,
Birmingham, 1999}, R.~Weikard and G.~Weinstein, eds., 271-282
Amer. Math. Soc./Internat. Press (2000).  arXiv math-ph/9910033,
mp\_arc 99-401.

\bibitem{LSYdoeb} E.H. Lieb, R. Seiringer, J. Yngvason, \textit{The
Ground State Energy and Density of Interacting Bosons in a Trap}, in
\textit{ Quantum Theory and Symmetries}, Goslar, 1999, H.-D.~Doebner,
V.K.~Dobrev, J.-D.~Hennig and W. Luecke, eds., pp. 101-110, World
Scientific (2000). arXiv math-ph/9911026, mp\_arc 99-439.

\bibitem{TRAP} W. Ketterle, N. J. van Druten, {\it Evaporative Cooling of
Trapped Atoms}, in B. Bederson, H.
Walther, eds.,  Advances in Atomic, Molecular and Optical Physics, {\bf
37}, 181--236, Academic Press (1996).

\bibitem{DGPS}
F. Dalfovo, S.\ Giorgini, L.P.\ Pitaevskii, and S.\ Stringari, {\it Theory of
Bose-Einstein condensation in trapped gases},
Rev. Mod. Phys. \textbf{71},
463--512 (1999).

\bibitem{S1999}
R. Seiringer, Diplom thesis, University of Vienna, (1999).

\bibitem{BO} N.N. Bogolubov,  J.  Phys. (U.S.S.R.)  {\bf 11}, 23
(1947); N.N. Bogolubov and D.N.  Zubarev,  Sov. Phys.-JETP {\bf 1}, 83
(1955).


\bibitem{Lee-Huang-YangEtc}K.~Huang, C.N.~Yang, Phys. Rev. {\bf
105}, 767-775
(1957); T.D.~Lee, K.~Huang, and C.N.~Yang,  Phys. Rev. {\bf 106},
1135-1145 (1957); K.A. Brueckner, K. Sawada, Phys. Rev. {\bf 106},
1117-1127, 1128-1135 (1957).; S.T. Beliaev, Sov. Phys.-JETP {\bf 7},
299-307 (1958); T.T. Wu, Phys. Rev. {\bf 115}, 1390 (1959);
N. Hugenholtz, D. Pines, Phys. Rev. {\bf 116}, 489 (1959); M.
Girardeau, R. Arnowitt, Phys. Rev. {\bf 113},
755 (1959); T.D. Lee,  C.N. Yang, Phys. Rev. {\bf 117}, 12 (1960).
 %%%%%%%%%%%%%%%%%%%%%%%%%%%%%%%%%%%%%%%%%%%%%%%%%%%%%%%%%%%%%%%%%%
\bibitem{Lieb63} E.H. Lieb, \textit{Simplified Approach to the Ground
State Energy of an Imperfect Bose Gas}, 
Phys. Rev. \textbf{130} (1963), 2518--2528. See also Phys. Rev. \textbf{133}
(1964), A899--A906 (with A.Y. Sakakura) and Phys. Rev. \textbf{134}
(1964), A312--A315 (with W. Liniger).

\bibitem{EL2} E.H.~Lieb, \textit{The Bose fluid}, in W.E.~Brittin, ed.,
Lecture Notes in
Theoretical Physics VIIC,  Univ. of Colorado Press,
pp.\ 175--224 (1964).

\bibitem{dyson}
F.J. Dyson, \textit{Ground-State Energy of a Hard-Sphere Gas},
Phys. Rev. \textbf{106}, 20--24 (1957).

\bibitem{BA} B. Baumgartner, \textit{The Existence of Many-particle Bound
States Despite a Pair Interaction with Positive Scattering Length},
J. Phys. A \textbf{30} (1997), L741--L747.


\bibitem{LL} E.H. Lieb, W. Liniger, \textit{Exact Analysis of an
Interacting Bose Gas. I.  The General Solution and the Ground State},
Phys. Rev. \textbf{ 130}, 1605-1616 (1963); E.H. Lieb, \textit{Exact
Analysis of an Interacting Bose Gas. II. The Excitation Spectrum},
Phys. Rev. \textbf{ 130}, 1616-1624 (1963).

\bibitem{TE} G. Temple, \textit{The theory of Rayleigh's Principle as
Applied to Continuous Systems},
Proc. Roy. Soc. London A \textbf{119} (1928),  276--293.

\bibitem{schick} M. Schick, \textit{ Two-Dimensional System of Hard Core
Bosons}, Phys. Rev.  A \textbf{ 3}, 1067-1073 (1971).

\bibitem{KLS} T. Kennedy, E.H. Lieb, S. Shastry,  The $XY$ Model has
Long-Range Order for all Spins and all Dimensions Greater than One, Phys.
Rev. Lett. textbf{ 61}, 2582-2584 (1988).


\bibitem{LN} E.H. Lieb, H. Narnhofer,  \textit{The Thermodynamic
Limit for Jellium}, J.  Stat. Phys. \textbf{ 12}, 291-310 (1975).
Errata J. Stat. Phys. \textbf{ 14}, 465 (1976).

\bibitem{FO} L.L. Foldy, \textit{Charged Boson Gas}, Phys. Rev.
\textbf{124}, 649-651 (1961); Errata \textit{ibid} \textbf{125}, 
2208 (1962). 

\bibitem{D2} F.J. Dyson, \textit{Ground State Energy of a Finite System
of Charged Particles}, J. Math. Phys. \textbf{8}, 1538-1545 (1967).

\bibitem{CLY} J. Conlon, E.H. Lieb, H-T. Yau \textit{The $N^{7/5}$ Law
for Charged Bosons}, Commun. Math. Phys. \textbf{ 116}, 417-448 (1988).

\bibitem{GS} G.M. Graf, J.P. Solovej, \textit{A correlation estimate
with applications to quantum systems with Coulomb interactions},
Rev. Math. Phys., 6 (No. 5a, Special Issue) 977-997 (1994).
mp\_arc 93-60.

\end{thebibliography}

\end{document}